\documentclass{mem}
\usepackage{natbib}\usepackage{txfonts}\usepackage{balance}
\usepackage{graphicx}
\usepackage[a4paper]{hyperref}
\idline{79}{1}
\begin{document}
\def\teff{$T\rm_{eff }$}
\def\kms{$\mathrm {km s}^{-1}$}

\title{
GALEX Ultraviolet photometry of NGC 2420: searching for WDs
}

   \subtitle{}

\author{
C. \,De Martino\inst{1},
\,L. \, Bianchi\inst{2},
\, I. \, Pagano\inst{3},
\,J. \, Herald\inst{2}
\and  D. \, Thilker\inst{2}
          }
  \offprints{C. De Martino}

\institute{
$^1$ Universit\`{a} degli Studi di Catania --
Dipartimento di Fisica e Astronomia, Via Santa Sofia 64,
I-95123 Catania, Italy, \email{demartino@oact.inaf.it}\\
$^2$ Johns Hopkins University of Baltimore --
Department of Physics and Astronomy, 3400 N. Charles St., Baltimore
21218,USA\\
$^3$ Istituto Nazionale di Astrofisica --
Osservatorio Astrofisico di Catania, Via S.Sofia 78,
I-95123 Catania, Italy
}

\authorrunning{C. De Martino

}

\titlerunning{UV imaging of NGC 2420}

\abstract{
We present color-magnitude diagrams of the open cluster NGC 2420, obtained from {\em Galaxy Evolution Explorer} (GALEX) 
ultraviolet images in FUV and NUV bands and {\em Sloan Digital Sky Survey} (SDSS) {\em u,g,r,i,z} photometry. Our goal is to search for and characterize hot evolved stars and peculiar objects in this open cluster, as part of a larger project aimed to study a number of open clusters in the Milky Way with GALEX and ground-based data. 
}
\maketitle{}

\section{Introduction}
NGC 2420 is a metal poor ({Z=0.008}) old open cluster (2.0 Gyr) at a distance of 2.48 kpc, well studied in the visible range \citep{sharma06}. From isodensity curves, \citet{sharma06} estimate an almost spherical core and a cluster coronal region with radial extent of 10 arcmin.
We present preliminary results of our search for hot evolved objects using combined UV and visible photometric measurements
from GALEX and SDSS data. 
The FUV and NUV GALEX bands are very sensitive to the detection of  hot stellar objects, and the combination of FUV to NIR bands allows us to uncover hot evolved objects in binary systems where the unevolved cooler companion dominates the optical light \citep{bianchi07}.

\section{GALEX and SDSS data}
The FUV and NUV photometry was measured from a GALEX observation made for the All-sky Imaging Survey (AIS)
 on Jan 04 2006 with an exposure time of 128~s (equal for the FUV and NUV images). The data were downloaded from the MAST archive.
GALEX provides simultaneous imaging in two UV bands, far-ultraviolet (FUV; {$\lambda_{\rm eff}$=1516 $\AA$}) 
and near-ultraviolet (NUV; {$\lambda_{\rm eff}$=2267 $\AA$}), with a circular field of view of 1.2$^o$ diameter \citep{martin05}.
 The GALEX pipeline photometry was not used because it merges nearby objects in the crowded parts of the cluster.
 Therefore, we performed custom photometry using DAOPHOT, for all sources detected on a 2-pxl-smoothed NUV image 
with significance above 5$\sigma$.
We used aperture photometry to extract the NUV and FUV magnitudes in the AB magnitude system, 
adopting an aperture of 6 arcsec. We applied aperture corrections following \citet{morrissey07}.

The optical data used in our analysis are taken from the Sloan Digital Sky Survey archive \citep{york00}. 
The SDSS photometric system covers from 300 nm to 1100 nm with five broad optical bands: u ({\em 3200-3800 $\AA$}), g ({\em 4100-5500 $\AA$}), r ({\em 5550-6950 $\AA$}), i ({\em 6950-8450 $\AA$}) and z ({\em 8500-9700 $\AA$}) \citep{fukugita04}.

\section{The source catalogue}
We matched the UV sources to pointlike optical sources using a match radius of 2.5 arcsec.
From the matched sources catalogue we extracted the objects within a 10~arcmin circular region 
centered at the NGC 2420 center (RA=07h38m28s, DEC=+21deg34m01s, J2000). 
The resulting catalog includes 344 NUV sources, 17 of which have no significant FUV detection. 
For comparison purposes, we extracted a sample of targets representing field stars from a region
 centered at RA=07h36m48s, DEC=+21deg14m06s (J2000) and covering an area equal to that of our cluster sample.

\section{Analysis}
We use the matched GALEX-SDSS photometry to build color-magnitude diagrams (e.g., Fig. 1) and 
color-color diagrams (examples in Fig.s 2 and 3) to investigate the cluster's stellar population. 
For the analysis, we further limit the sample to objects with good photometry. A detailed
discussion will be provided elsewhere. An isochrone (from Girardi et al. 2004) for a 2Gr old population
is shown on the CMD.  There is clearly a concentration of objects along the isochrone, presumably cluster
members, in addition to a sparse distribution of sources matching the distribution of the field stars. 
Model colors (from Bianchi et al. 2007) for stars of varying T$_{eff}$ and gravity values down to log~g=9,
reddened with  E(B-V)=0.04~mag, are shown in the color-color diagrams. 
Two sources in the NGC 2420 area appear to be hot evolved objects, from their photometric colors. 
In Fig. 4 we plot the spectral energy distribution  (SED) of star A, 
together with two pure-H models, computed with the TLUSTY code. Assuming that star-A is at the cluster's distance,
its luminosity would require a small radius, making it a candidate  evolved object. However, we point
out that a similar small number of hot evolved star candidates is found in the control sample (field stars),
and follow-up data are needed to establish cluster membership of the hot stars. 

\begin{figure}[h!]
\includegraphics[width=6.5cm,height=5.85cm]{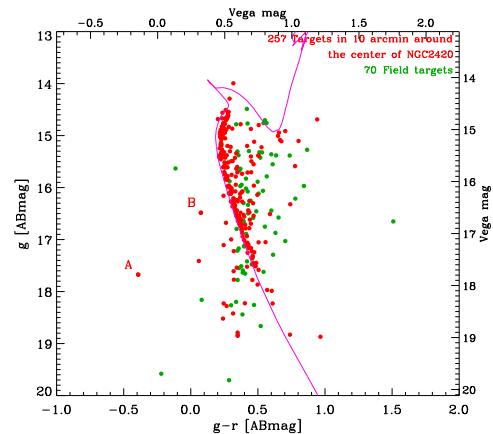}
\caption{\footnotesize
\em The [g] vs [g-r] HR diagram for the UV sources in the NGC2420 area (red dots) and in the control field (green dots). 
The continuous line represents a theoretical isochrone from the evolutionary models of \citet{gira04}, with age of 2 Gyr and 
metallicity  z=0.008.
}
\label{Figura 1}
\end{figure}

\begin{figure}[h!]
\includegraphics[width=6.5cm,height=5.85cm]{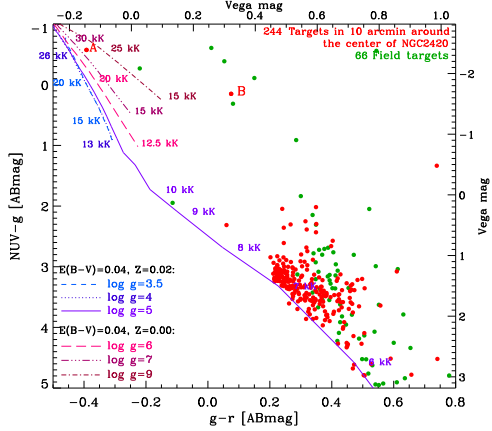}
\caption{\footnotesize
\em The [NUV-g] vs [g-r] color-color diagram. Model colors (described in Bianchi et al. 2007) 
are shown with different symbols and colors, explained in the legends, for a 
reddening of E(B-V)=0.04. The model colors are obtained from Kurucz spectra (solar metallicity) for 
gravities of log~g=5 and lower, and from TLUSTY models (pure-H) for high gravity stars.}
\label{Figura 2}
\end{figure}

\begin{figure}[h!]
\includegraphics[width=6.5cm,height=5.85cm]{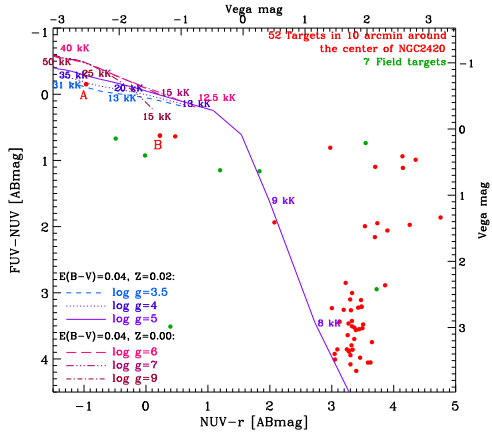}
\caption{\footnotesize
\em The [FUV-NUV] vs [NUV-r] color-color diagram. Symbols as in Figure 2.
}
\label{Figura 3}
\end{figure}

\begin{figure}[h!]
\includegraphics[width=6.5cm,height=5.85cm]{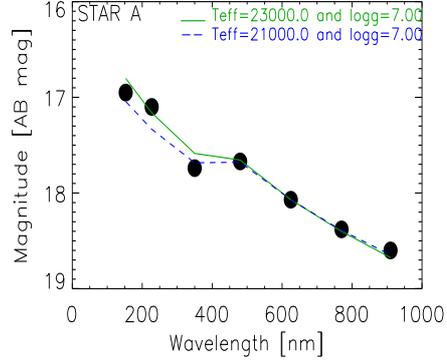}
\caption{\footnotesize
\em SED (FUV to near-IR) of a hot star in the NGC2420 area. 
Black dots are the measured magnitudes,  plotted at the effective
wavelength of each filter. Photometric errors are smaller than the dots. 
Two pure-H TLUSTY models are shown, scaled to the measured magnitude in the r-band. At the distance of 2.48kpc, the scaling factor would imply R=0.1 R$_{\rm \odot}$ (for log g =7) for the UV source.
}
\label{Figura 4}
\end{figure}

\section{Future work}
SED fitting of the photometry with model colours is in progress for all measured sources, to determine their physical parameters.
Further work includes the study of other MW clusters, and follow-up spectroscopy of the hot evolved candidate objects.

\begin{acknowledgements}
Based on archival data from the NASA Galaxy Evolution Explorer (GALEX) which is operated 
for NASA by the California Institute of Technology under NASA contract NAS5-98034.  
LB, DT and JH acknowledge support from the GALEX project 
and from NASA grant NNX07AJ47G (GI cycle 3). 
Some of the data presented in this paper were obtained from the Multimission Archive at the Space Telescope Science Institute (MAST). STScI is operated by the Association of Universities for Research in Astronomy, Inc., under NASA contract NAS5-26555. Support for MAST for non-HST data is provided by the NASA Office of Space Science via grant NAG5-7584 and by other grants and contracts.
\end{acknowledgements}

\bibliographystyle{aa}

\end{document}